\documentclass[amssymb,aps,twocolumn,floats,showpacs]{revtex4-1}
\usepackage{color,graphicx,pstricks,float}
\usepackage{natbib}

\usepackage{fancybox}
\usepackage{epsfig}
\usepackage{graphicx}
\usepackage{tabularx}
\usepackage{fancybox}
\usepackage{epsfig}
\usepackage{amsmath}
\begin{document}

\title{Topological Transitions in a Model for Proximity Induced Superconductivity}

\author{Navketan Batra}
\author{Swagatam Nayak}
\author{Sanjeev Kumar}

\address{
Indian Institute of Science Education and Research (IISER) Mohali, Sector 81, S.A.S. Nagar, Manauli PO 140306, India \\
}

\begin{abstract}

Using a prototype model for proximity induced superconductivity on a bilayer square lattice, we show that interlayer tunneling can drive change in topology of the Bogoliubov quasiparticle bands. Starting with topologically trivial superconductors, transitions to a non-trivial $p_x + {\rm i} p_y$ state and back to another trivial state are discovered.
We characterize these phases in terms of edge-state spectra and Chern indices. We show that these transitions can also be controlled by experimentally viable control parameters, the bandwidth of the metallic layer and the gate potential. Insights from our results on a simple model for proximity induced superconductivity may open up a new route to discover topological superconductors.
\end{abstract}
\date{\today}


\maketitle

\noindent
\underline{\it Introduction:}
Last decade witnessed a paradigm shift in the general approach to understand electronic properties of crystalline solids. 
Knowledge of topological character of the single particle bands turns out to be crucial for comprehending certain exotic electronic properties \cite{Kruthoff2017}.
This change in approach originated in the discovery of topological insulators, materials that are insulating in the bulk but support topologically-protected metallic surface states  \cite{Kane2005c, Kane2005d, Hasan2010, Moore2010}.
Superconductivity, a fascinating phenomenon in its own right, has intrigued physicists time and again by appearing in unexpected settings. The most recent examples are the `magic angle' superconductivity in bilayer graphene \cite{Cao2018, Wu2018}, and the tip-induced superconductivity in Cd$_3$As$_2$ \cite{Aggarwal2016, Wang2016}. The discovery of topological insulators motivated a similar search for materials that are superconducting (SC) in the bulk, but support gapless modes on surfaces \cite{Qi2011, Bansil2016, Sato2017, Kallin2016a, Fu2008}. The key is to find ways to alter the band structure of the relevant Bogoliubov quasiparticle bands. In addition to being of fundamental interest, topological superconductors are considered as building blocks of decoherence-free quantum computers \cite{Nayak2008, Potter2011, Beenakker2013, Nadj-Perge2014, Ruby2015, Beenakker2016}.

The existence of superconductivity in atomically thin layers has recently been reported by various groups. 
Superconductivity in a single-atomic layer film of Pb grown on Si(111) substrate was observed \cite{Zhang2010}.
A monolayer of CuO$_2$ grown on cuprate substrate was found SC \cite{Zhu2016}. 
Unconventional, possibly topological, superconductivity is reported at the interface between LaAlO$_3$ and SrTiO$_3$ \cite{Scheurer2015}.
Proximity induced topological superconductivity has been proposed for bilayer graphene \cite{Alsharari2018}. 
Superconductivity can be induced, with no accompanying structural changes, in NbAs$_2$ by applying external pressure \cite{Li2018}.
The tip-induced SC phase of Cd$_3$As$_2$ has recently been stabilized in thin films \cite{Suslov2018}.
These diverse material examples share a common feature -- superconductivity appears when coupling between two layers is altered. 
Motivated by the role of proximity effect in a variety of superconductors, we explore this effect
in a general setting with focus on inducing topologically nontrivial character in superconductors.

In this work, we show that simple interlayer tunneling can alter the topology of Bogoliubov quasiparticle bands. This is achieved in a prototypical model of proximity induced superconductivity where a SC layer of square lattice is tunneling-coupled to a tight-binding layer. 
The calculations are performed within an unrestricted mean-field approach, allowing for existence of multiple symmetries of the SC order parameters (OPs). 
We find that interlayer tunneling can induce transition to topologically non-trivial state with $p_x + {\rm i} p_y$ symmetry. 
A complete characterization of the band topology is carried out via Berry curvature and Chern number calculations complemented by the edge spectra in cylinder geometry.
An interplay among different OP symmetries in a two band setting is responsible for the transitions. Additionally, we find a connection between the topological transitions and the Lifshitz transitions in the underlying metallic bands.
The generic nature of the model suggests that this can be applicable, with suitable variations, to a wide class of systems that invoke proximity effect \cite{Cao2018,Aggarwal2016,Wang2016,Zhang2010,Zhu2016,Scheurer2015,Alsharari2018,Li2018}.

\noindent
\underline{\it Bilayer Model for Proximity Induced Superconductivity:} 
As a prototype model for proximity induced superconductivity, we consider an Extended Attractive Hubbard Hamiltonian (EAHM) defined on a 2D square lattice coupled via inter-layer tunneling to a tight-binding layer. 
The model is described by the Hamiltonian,

\begin{eqnarray}
H & = & H_1 + H_2 + H_{12}, \nonumber  \\
H_{1} & = &  - t_1 \sum_{\langle ij\rangle,\sigma} [ c^\dagger_{i \sigma 1} c^{}_{j \sigma 1} + H.c.] -
\mu_1 \sum_{i \sigma}c^\dagger_{i \sigma 1} c^{}_{i \sigma 1} \nonumber \\
& & -U \sum_{i} n_{i \uparrow 1} n_{i \downarrow 1} - V \sum_{\langle ij \rangle} n_{i 1}n_{j 1}, \nonumber \\
H_{2} & = &  - t_2 \sum_{\langle ij\rangle,\sigma} [ c^\dagger_{i \sigma 2} c^{}_{j \sigma 2} + H.c.] -
\mu_2 \sum_{i \sigma}c^\dagger_{i \sigma 2} c^{}_{i \sigma 2}, \nonumber \\
H_{12} & = & -{\tilde t} \sum _{i \sigma} [ c^\dagger_{i \sigma 1} c^{}_{i \sigma 2} + H.c.].
\label{Ham}
\end{eqnarray}
\noindent
Here $c_{i \sigma l} (c_{ i\sigma l}^\dagger$) annihilates (creates) an electron in layer $l$
at site ${i}$ with spin $\sigma$, $\langle ij \rangle$ implies that sites
$i$ and $j$ are nearest neighbors within a layer. $\mu_l$ is the layer-dependent chemical
potential, with $\mu_1 - \mu_2$ being equivalent to gate potential.
The layer-resolved local number operators are given by
$n_{i \sigma l} = c_{i\sigma l }^{\dagger}c^{}_{i\sigma l}$, and $n_{i l} = n_{i \uparrow l } + n_{i \downarrow l}$. $U$($V$) denotes the strength of
on-site (nearest neighbor) attractive interaction in layer 1.
Using $t_1=1$ as the basic energy scale, and restricting ourselves to zero temperatures ($T=0$), we are left with six independent parameters in the Hamiltonian, {\it viz.}, $t_2$, ${\tilde t}$, $U$, $V$, $\mu_1$ and $\mu_2$. In order to avoid a brute force exploration of this large parameter space, we make use of the recently reported comprehensive phase diagram of the monolayer model \cite{Nayak2018}. We set $U = 1$ throughout the paper.

\noindent
\underline{\it Method:} We analyze the Hamiltonian in Eq.~(\ref{Ham}) by making a mean-field Bogoliubov-deGennes (BdG) approximation, for 
the interaction term \cite{deGennes1999, Jian-xin2016}.
In the intersite attractive term we ignore the same-spin attraction parts $n_{i \uparrow} n_{j \uparrow}$ and $n_{i \downarrow} n_{j \downarrow}$ \cite{Nayak2018}.
Following the standard mean-field decoupling in the pairing channel, we arrive at the pairing Hamiltonian for layer 1,

\begin{eqnarray}
H^{BdG}_{1} & = & - t_1 \sum_{\langle ij \rangle,\sigma} \left[ c_{i\sigma 1}^\dagger
c_{j\sigma 1} + H.c. \right] - \mu_1 \sum_{i \sigma}c^\dagger_{i \sigma 1} c^{}_{i \sigma 1} \nonumber \\
& & - U \sum_{i}\left[\Delta_{i, 1}c_{i\uparrow 1}^\dagger c_{i\downarrow 1}^\dagger + H.c. \right] \nonumber \\
& & - V \sum_{i \gamma} \left[\Delta^{+}_{i, \gamma, 1}c_{i\uparrow 1}^\dagger c_{i+\gamma \downarrow 1}^\dagger + \Delta^{-}_{i, \gamma, 1}c_{i-\gamma \downarrow 1}^\dagger c_{i \uparrow 1}^\dagger
+ H.c. \right] \nonumber \\
& & +U \sum_i | \Delta_{i, 1} |^2 + V \sum_{i \gamma} \left[ |\Delta^+_{i, \gamma, 1}|^2 + |\Delta^-_{i, \gamma, 1}|^2  \right].
\label{BdGequation}
\end{eqnarray}        
\noindent
In the above we have introduced the pair expectation values in the ground state as, $\Delta_{i,l} = \langle c_{i\downarrow l} c_{i\uparrow l} \rangle$, $\Delta^+_{i,\gamma,l} = \langle c_{i+\gamma \downarrow l} c_{i \uparrow l} \rangle$, and 
$\Delta^-_{i,\gamma,l} = \langle c_{i-\gamma \downarrow l} c_{i \uparrow l} \rangle$, where $\gamma$ denotes the unit vectors $+\hat {\bf x}$ and $+\hat {\bf y}$ on the square lattice, and $l$ is the layer index.
Note that we do not impose the commonly used spin-singlet symmetry constraint on the pair expectation values. In general, $\Delta^+_{i, \gamma, l} \neq \Delta^-_{i+\gamma, \gamma, l}$, and spin-triplet component of superconductivity is allowed to exist as a broken-symmetry mean-field phase. 
Indeed, it has recently been shown that a triplet SC state with $S_z = 0$ is possible in models and experiments \cite{Diesch2018, Nayak2018}.

Assuming translational invariance, $\Delta_{i,l} \equiv \Delta_{0,l}$ and $\Delta^{\pm}_{i,x/y,l} \equiv \Delta^{\pm}_{x/y,l}$, we work in Fourier space by using, $c_{i \sigma l} = N_s^{-1/2} \sum_{\bf k} e^{-{\rm i} {\bf k} \cdot {\bf r}_i} c_{{\bf k} \sigma l}$ and $c^{\dagger}_{i \sigma l} = N_s^{-1/2} \sum_{\bf k} e^{{\rm i} {\bf k} \cdot {\bf r}_i} c^{\dagger}_{{\bf k} \sigma l}$, $N_s$ being the number of sites in each layer. Up to a constant, the resulting mean-field Hamiltonian is given by,

\begin{eqnarray}
H^{MF} & = & \sum_{\bf k} \left( \sum_{\sigma,l} \xi_l({\bf k}) c^{\dagger}_{{\bf k}\sigma l}c_{{\bf k}\sigma l} + \left[ \Delta^{\uparrow\downarrow}_1({\bf k}) c^{\dagger}_{{\bf k}\uparrow 1}c^{\dagger}_{{\bf -k}\downarrow 1} + H.c. \right] \right. \nonumber \\ 
& & \left. - \tilde{t} \sum_{\sigma} \left[c^{\dagger}_{{\bf k}\sigma 1}c^{}_{{\bf k}\sigma 2} +  c^{\dagger}_{{\bf k}\sigma 2}c^{}_{{\bf k}\sigma 1}\right] \right),
\label{Ham-mf}  
\end{eqnarray}

\noindent
where, 

\begin{eqnarray}
\xi_{l}({\bf k}) & = & -2t_l(\cos{k_x}+\cos k_y)-\mu_l \nonumber \\
\Delta^{\uparrow\downarrow}_1({\bf k}) & = & -U\Delta_{0,1} - V (e^{-ik_x}\Delta^{+}_{x,1} + e^{ik_x}\Delta^{-}_{x,1} \nonumber \\ 
& & + e^{-ik_y}\Delta^{+}_{y,1} + e^{ik_y}\Delta^{-}_{y,1})
\end{eqnarray}
\noindent
The Hamiltonian Eq. (\ref{Ham-mf}) is diagonalized using bilayer generalization of the Bogoliubov transformations, $c_{{\bf k}\sigma l} = \sum_n \big( u_{{\bf k}\sigma l}^n\gamma_n - \sigma v_{{\bf k}\sigma l}^{n^*}\gamma^{\dagger}_n \big)$ \cite{deGennes1999, SM}. 
Following the standard Bardeen-Cooper-Schrieffer (BCS) approach, the groundstate is constructed as a vacuum of Bogoliubov quasiparticle.
The mean-field variables are computed in terms of the transformation coefficients and the Fermi factors, via,
\begin{eqnarray}
\Delta_{0,l} & = & \frac{1}{N_s} \sum_{{\bf k}, n} u^{n}_{{\bf k}\uparrow l} v^{n*}_{{\bf -k}\downarrow l} f(E_n) \nonumber \\
\Delta^{\pm}_{x/y,l} & = & \frac{1}{N_s} \sum_{{\bf k}, n} e^{\mp ik_{x/y}} u^{n}_{{\bf k}\uparrow l} v^{n*}_{{\bf -k}\downarrow l} f(E_n).
\end{eqnarray} 
\noindent The averages, $\Delta_{0,1}$ and $\Delta^{\pm}_{x/y,1}$, are calculated self-consistently with convergence criterion set to $10^{-5}$. The standard SC OPs are defined in terms of the converged parameters as,
\begin{eqnarray}
 \Delta^l_s & = & \Delta_{0,l}  \nonumber \\
 \Delta^l_{d/s^*} & = & [(\Delta^+_{x, l} + \Delta^-_{x, l}) \mp (\Delta^+_{y, l} + \Delta^-_{y, l})]/4 \nonumber \\
 \Delta^l_{p_{x/y}} & = & [\Delta^+_{x/y, l} - \Delta^-_{x/y, l}]/2.
\label{OP-def}
\end{eqnarray}

\noindent
The $s$-, $s^*$-, $p$- and $d$-wave OPs defined above have their usual meaning as can be verified from the ${\bf k}$-dependence of $\Delta^{\uparrow\downarrow}_1({\bf k})$ in limiting cases \cite{Nayak2018}. 

\begin{figure}[t!]
\includegraphics[width=.98 \columnwidth,angle=0,clip=true]{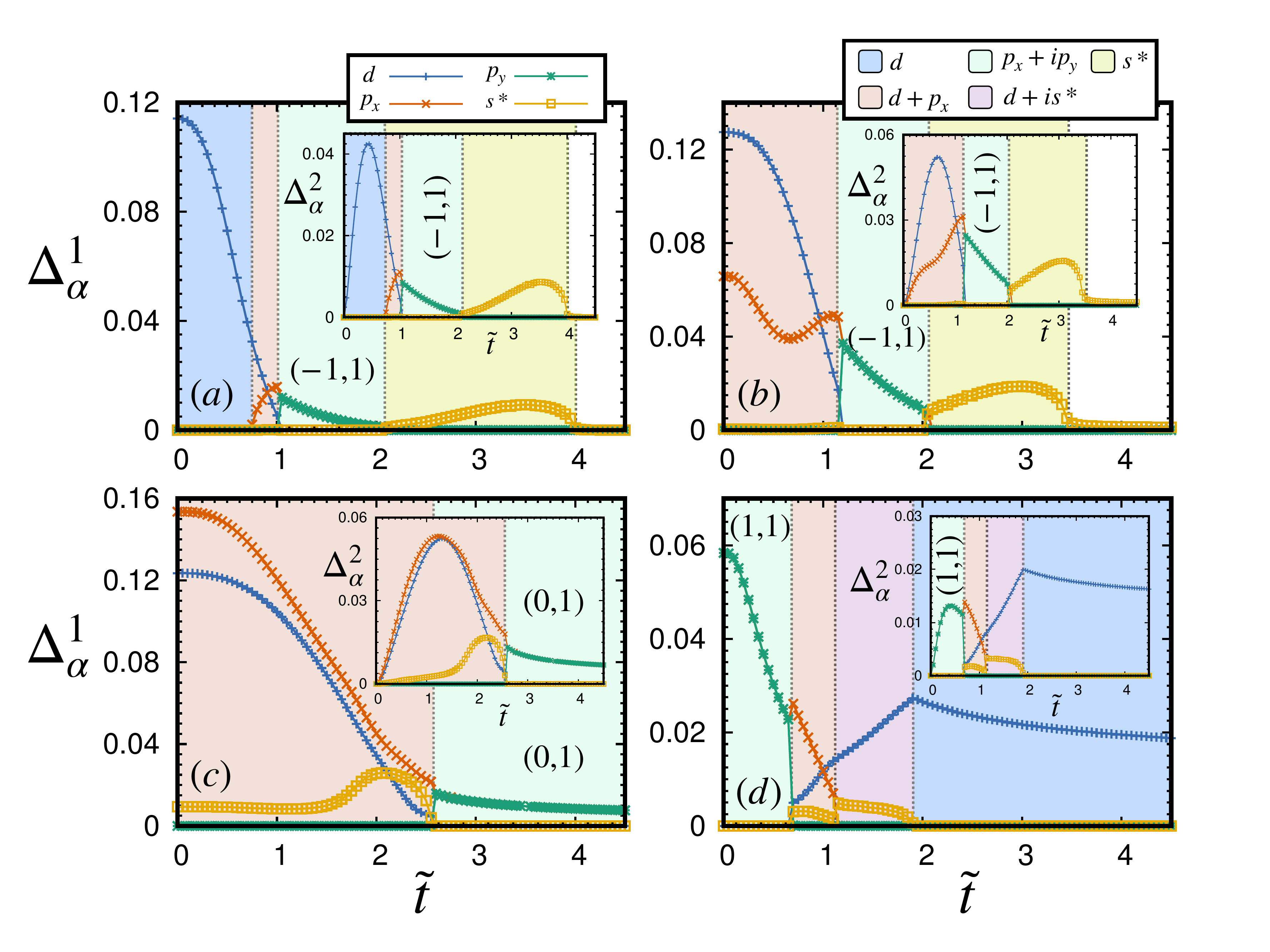}
\caption{(Color online) ($a$)-($d$) The self-consistent values of various OPs as a function of interlayer hopping $\tilde{t}$ for $U=1$ and $t_2 = 1$.
Insets show the OP variations in the proximity layer. Different background shades correspond to qualitatively different OP symmetries as indicated. The choice of other parameters is: ($a$) $V=1.8$, $ \langle n_1 + n_2 \rangle = 2.0$, ($b$) $V=2.5$, $ \langle n_1 + n_2 \rangle = 1.8$, ($c$) $V=3.5$, $ \langle n_1 + n_2 \rangle = 1.4$, and ($d$) $V=2.2$, $ \langle n_1 + n_2 \rangle = 0.9$. Integer pairs denote the Chern indices for the two bands wherever at least one is non-zero.
}
\label{fig1}
\end{figure}

\noindent
\underline{\it Tunneling driven transitions:} 
We begin by presenting in Fig. \ref{fig1} the effect of interlayer hopping $\tilde{t}$ on the SC OPs. We select model parameters such that in the decoupled limit, $\tilde{t} = 0$, different OP symmetries are realized in the SC layer \cite{Nayak2018}. In Fig. \ref{fig1} ($a$), we begin with a $d$-wave OP in the SC layer. For an arbitrarily small value of inter-layer hopping, the $d$-wave OP is induced in the second layer. Near $\tilde{t} = 0.8$, a $p$-wave component appears in the solution for both the layers. At $\tilde{t} = 1$, the $d$-wave OP vanishes and the stable solution acquires the $p_x + {\rm i} p_y$ form. This unconventional OP reduces gradually upon increasing $\tilde{t}$, and near $\tilde{t} = 2.2$ another transition to an extended $s$-wave, $s^*$, form occurs. Therefore, allowing for broken symmetry phases at the mean field level, we find multiple transitions tuned by inter-layer hopping. These transitions are mirrored in the second layer via the proximity effect (see insets in Fig. \ref{fig1} ($a$)-($d$)). Eventually, beyond a critical value of the interlayer hopping superconductivity ceases to exist in both the layers in agreement with the previous report on proximity effect \cite{Zujev2014}.

In Fig. \ref{fig1} ($b$) we demonstrate the occurrence of these transitions starting with a $d+p_x$-wave OP in the SC layer. The sequence of change in OP symmetries is $d+p_x$ to $p_x + {\rm i} p_y$ to extended $s$. The sequence of transitions can be reversed if we begin with a $d+p_x+s^*$ or $p_x + {\rm i} p_y$ state, as shown in Fig. \ref{fig1} ($c$)-($d$). We find that some of these transitions are associated with Lifshitz transitions tuned by $\tilde{t}$ in the non-interacting Hamiltonian
\cite{SM, Volovik2017, Chen2012, Lin2010, Shi2018}. 
Note that the $p_x + {\rm i} p_y$ form of the OP can lead to different band topology depending on the other model parameters (compare Chern indices in panels (a) and (c) in Fig. \ref{fig1}). 

In these calculations we have kept $\mu_2 = \mu_1$ and the densities in the two layers are allowed to be different. We find that the densities in two layers become very close to each other with increasing $\tilde{t}$ \cite{SM}. We also perform calculations enforcing equal density in the two layers, which generally requires $\mu_2 \neq \mu_1$. The results are qualitatively identical to those discussed in Fig. \ref{fig1} \cite{SM}. 

\noindent
\underline{\it Characterizing the Topologically Nontrivial Phases:} 
We follow two standard approaches to characterize the nontrivial SC states. The first approach involves analyzing Berry curvature and computing topological invariants, known as Chern numbers, associated with each quasiparticle band. 
We employ an efficient method to calculate Chern numbers, in the discrete Brillouin zone, by making use of $U(1)$ link variable \cite{Fukui2005},

\begin{figure}[t!]
\includegraphics[width=.98 \columnwidth,angle=0,clip=true]{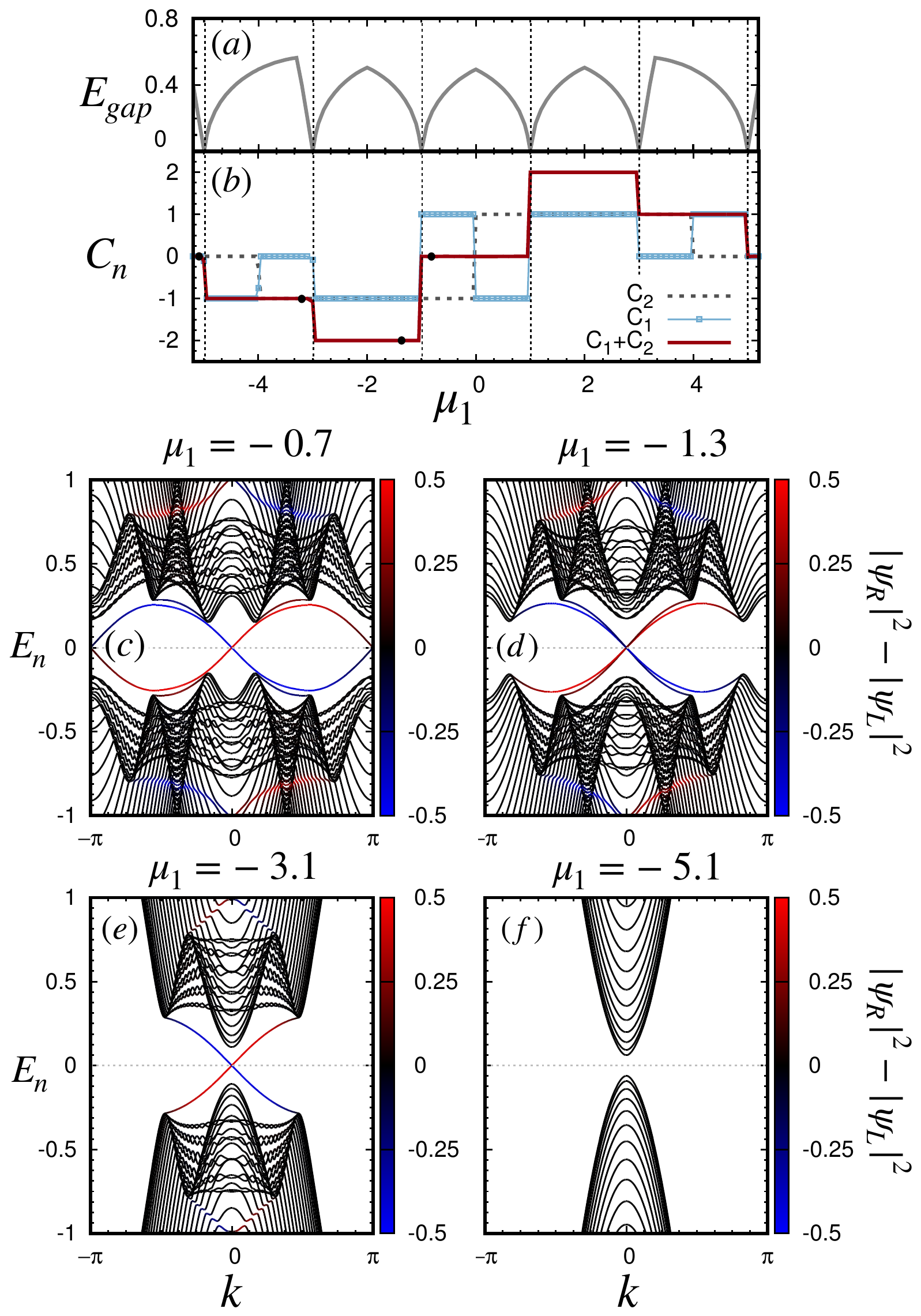}
\caption{(Color online) ($a$) Variation of the SC gap with chemical potential for $p_x + {\rm i} p_y$ state. ($b$) Chern numbers for each band and total Chern index for the same state as in (a). Note that every jump in total Chern index coincides with a closing and reopening of the SC gap. (c)-(f) Edge state spectrum in the cylinder geometry for different values of $\mu_1 = \mu_2$. We keep $t_2 = {\tilde t} = 1$ for all the results shown here. 
}
\label{fig2}
\end{figure}

\begin{eqnarray}
U_{n}^{\hat{\epsilon}}({\bf k}) = \frac{\langle n({\bf k})  | n({\bf k+\hat{\epsilon}}) \rangle}{|\langle n({\bf k})  | n({\bf k+\hat{\epsilon}}) \rangle|}.
\end{eqnarray}

\noindent In the above, $\hat{\epsilon}$ is a vector connecting nearest neighbour points in the discrete Brillouin zone. The Berry curvature, which is gauge invariant, can be calculated as the total phase along a closed loop as,

\begin{eqnarray}
F_n({\bf k}) = \frac{1}{i}\ln{U_n^{\bf \hat{x}}({\bf k}) U_n^{\bf \hat{y}}({\bf k + \hat{x}}) U_n^{{\bf -\hat{x}}}({\bf k + \hat{x}+\hat{y}}) U_n^{\bf -\hat{y}}({\bf k+\hat{y} }) }.
\nonumber \\
\end{eqnarray}

\noindent
Note that the Berry curvature is defined within the principle branch of the logarithm, $ -\pi < F_n({\bf k}) \leq \pi$. Summing it over the Brillouin zone gives $2 \pi C_n$, where $C_n$ is the Chern number for the $n^{\text{th}}$ band.

To illustrate how the topological character of a SC state changes, we select the $p_x + {\rm i} p_y$ form of the OP. In Fig. \ref{fig2}(a), we show the SC gap as a function of chemical potential. The gap closes and reopens upon varying $\mu_1$. Each such gap closing is associated with a change in the topological character of the bands. This is evident from Fig. \ref{fig2}(b), where we show band-specific as well as total Chern numbers.

The second approach is to compute the edge-state spectra by imposing open boundary conditions in one of the directions, leading to cylinder geometry, and plotting the tower of states as a function of $k_x$ or $k_y$. Note that only one of $k_x$ and $k_y$ is a good quantum number in the cylinder geometry.
The edge-state spectra are shown in Fig. \ref{fig2}(c)-(f) for representative values of $\mu$.
The color code on the energy eigenvalues represents the difference of the weight on left and that on right edges of the corresponding state.
Fig. \ref{fig2}(c) shows a case where two states cross the bulk gap, however the gap is crossed twice. Although both bands have a nonzero ($\pm 1$) Chern index, the total Chern index is zero. For $\mu = -1.1$, gap opens close to the Brillouin zone boundary and both the edge states cross the gap only once. In this case the Chern numbers for the two bands add up leading to a total Chern number of $-2$. For $\mu = -3.1$, one of the bands pulls away and only one pair of edge states remain. The corresponding Chern number for one of the bands becomes zero, leading to a total Chern number of -1.

Interestingly, these transitions can also be viewed as Lifshitz transitions in the emergent one-dimensional metallic system residing on the edges in the cylinder geometry. The Fermi surface of this one-dimensional metallic system consists of discrete points. Each instance of change in total Chern number is accompanied by disappearance of a point from the zero dimensional Fermi surface, which can be viewed as a Lifshitz transition in one lower dimension.

\begin{figure}[t!]
\includegraphics[width=.99 \columnwidth,angle=0,clip=true]{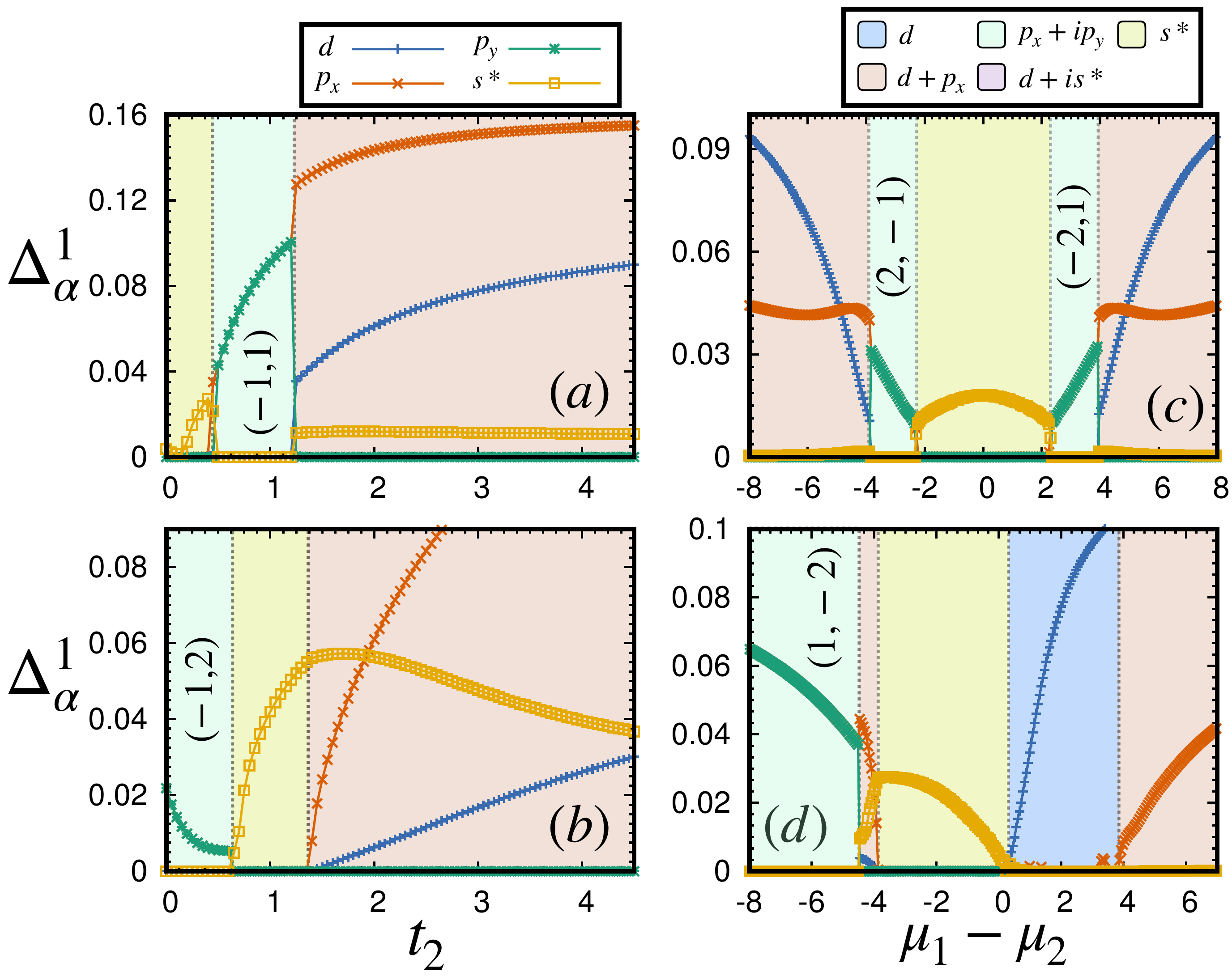}
\caption{(Color online) ($a$)-($b$) The values of various OPs as a function of metallic layer hopping $t_2$ for, ($a$) ${\tilde t} = 1.5$, $ \langle n_1 + n_2 \rangle = 1.8$ , $V=4$, and ($b$) ${\tilde t} = 2.5 $, $ \langle n_1 + n_2 \rangle = 1.6$ , $V=4$. ($c$)-($d$) Variations in the OPs with the gate potential $\mu_1 - \mu_2$ for, (c) $t_2 = 1.0$, ${\tilde t} = 2.6 $, $\mu_1 = 0$, $V = 2.5$ and (d) $t_2 = 1.0 $, ${\tilde t} = 2.25$, $\mu_1 = -1$, $V = 2.5$. Integer pairs denote the Chern indices for the two bands wherever at least one is non-zero.
}
\label{fig3}
\end{figure}

\noindent
\underline{\it Tuning the Transitions by Bandwidth and Gate Potential:}
Having shown the existence of unconventional phases in the bilayer model, we ask if one can tune the system across these transitions by using experimentally viable control parameters. To this end we present the effect of change in bandwidth of the metallic layer and that of gate potential on SC OPs. In Fig. \ref{fig3}($a$), we show the change in various OPs as a function of intralayer hopping $t_2$. Taking $t_2 = 1$ as a reference point, we find that a $p_x + {\rm i} p_y$ state can be tuned to $s^*$-wave ($d+p_x+s^*$-wave) state by decreasing (increasing) the bare bandwidth. The sequence of transitions can be altered, as we saw in case of transitions tuned by $\tilde{t}$, by selecting a different starting point (see Fig. \ref{fig3}($b$)).
From an experimental viewpoint the most easily tunable parameter in a possible realization of this model is the difference between on-site potentials in the two layers. We show that the nontrivial transitions in terms of OP symmetries discussed above can also be tuned with the help of gate potential $\mu_1 - \mu_2$. Two representative cases are shown in Fig. \ref{fig3}($c$)-($d$). 

A comment regarding the symmetry aspects of the SC solutions is in order. 
The various unusual SC states reported here are an outcome of a broken symmetry mean-field analysis. The stability of these phases is controlled by energetics, which relies crucially on the presence of two bands. The mixed symmetry phases, such as $d + p_x$, are examples of spontaneous breaking of parity symmetry, an unusual effect that has recently been observed in experiments \cite{Trenkwalder2016, Shu2012, Giacosa2010, Sergienko2004, Read2000}. In a realistic scenario, presence of additional symmetry breaking terms, such as Rashba coupling, is expected to further stabilize the unusual mixed symmetry states reported here \cite{Choy2011}.

\noindent
\underline{\it Conclusion:} 
We have shown that a prototype model of proximity induced superconductivity displays transitions between topologically trivial and nontrivial SC states. The model consists of an EAHM layer coupled to a tight-binding layer, both defined on a square lattice. A general treatment of the model that allows for various broken symmetry SC OPs is important to obtain the nontrivial topological transitions reported in this work. We characterize different phases in terms of their Chern indices and edge-state spectra. We further show that the transitions can be tuned by realistic parameters such as the gate potential and relative bandwidth of the two layers. Our results are directly relevant to systems that exhibit superconductivity in atomically thin layers, such as,
monolayer of CuO$_2$ \cite{Zhu2016}, bilyer graphene \cite{Alsharari2018}, NbAs$_2$ \cite{Li2018}, etc. Bringing such systems in proximity to a superconductor may lead to topological transitions via the change in form of SC OPs.
In our calculations, the topological transitions are present for moderate to strong values of attractive Hubbard parameters. Therefore, superconductors that are on the Bose-Einstein condensate (BEC) side of the BCS-BEC crossover are likely to host such effects. Possible candidates are the multiorbital superconductors Fe$_{1+y}$Se$_{1+x}$Te$_{1-x}$ which allow for a tuning across the BCS-BEC crossover \cite{Lubashevsky2012,Rinott2017}. 
These predictions can also be tested using ultracold Fermions on optical lattices where the interaction strength can be tuned with the help of Feshbach resonances \cite{Stewart2008, Chin2004, Partridge2005, Grass2016}.

\noindent
\underline{\it Acknowledgments:}
We acknowledge the use of High Performance Computing Facility at IISER Mohali.


\begin{center}{\bf Supplemental Material}
\end{center}

\noindent {\underline{\it Bogoliubov-de Gennes method for the bilayer}:}
In this section we derive the Bogoliubov-de Gennes (BdG) equations for the bilayer model. Once again, the bilayer mean field Hamiltonian is given by,

\begin{eqnarray}
H^{MF} & = & \sum_{\bf k} \Bigg( \sum_{\sigma,l}\xi_l({\bf k}) c^{\dagger}_{{\bf k}\sigma l} c_{{\bf k}\sigma l} + \left[ \Delta_{1}^{\uparrow\downarrow}({\bf k}) c^{\dagger}_{{\bf k}\uparrow 1} c^{\dagger}_{{\bf -k}\downarrow 1}  + H.c.\right] \nonumber \\
& & -\tilde{t}\sum_{\sigma} \left[ c^{\dagger}_{{\bf k}\sigma 1}c_{{\bf k}\sigma 2} + c^{\dagger}_{{\bf k}\sigma 2}c_{{\bf k}\sigma 1} \right] \Bigg),
\end{eqnarray} 

\noindent
where,

\begin{eqnarray}
\xi_{l}({\bf k}) & = & -2t_l(\cos{k_x}+\cos k_y)-\mu_l, \nonumber \\
\Delta^{\uparrow\downarrow}({\bf k}) & = & -U\Delta_0 - V (e^{-ik_x}\Delta^{+}_{x} + e^{ik_x}\Delta^{-}_{x}, \nonumber \\ 
& & + e^{-ik_y}\Delta^{+}_{y} + e^{ik_y}\Delta^{-}_{y}).
\end{eqnarray}

\noindent
The commutation relations of the mean-field Hamiltonian with the electronic creation and annihilation operators are given by,

\begin{eqnarray}\label{ecommutation}
\left[ c_{{\bf k}\uparrow 1},H^{MF} \right] &=& \xi_{1}({\bf k}) c_{{\bf k}\uparrow 1} + \Delta^{\uparrow \downarrow}({\bf k}) c^{\dagger}_{{\bf -k}\downarrow 1} - \tilde{t} c_{{\bf k}\uparrow 2}, \nonumber \\
\left[ c^{\dagger}_{{\bf -k}\downarrow 1},H^{MF} \right] &=& -\xi_{1}({\bf -k}) c_{{\bf -k}\downarrow 1} + \Delta^{\uparrow \downarrow^{*}}({\bf k}) c^{\dagger}_{{\bf k}\uparrow 1} + \tilde{t} c_{{\bf -k}\downarrow 2}, \nonumber \\
\left[ c_{{\bf k}\uparrow 2},H^{MF} \right] &=& \xi_{2}({\bf k}) c_{{\bf k}\uparrow 2} - \tilde{t} c_{{\bf k}\uparrow 1}, \nonumber \\
\left[ c^{\dagger}_{{\bf -k}\downarrow 2},H^{MF} \right] &=& -\xi_{2}({\bf -k}) c_{{\bf -k}\downarrow 2} + \tilde{t} c_{{\bf -k}\downarrow 1}.
\end{eqnarray}
\noindent Note that instead of writing the commutation relations using $\{c_{{\bf k}\uparrow l},c_{{\bf k}\downarrow l}\}$ we write them it terms of $\{c_{{\bf k}\uparrow l},c^{\dagger}_{{\bf -k}\downarrow l}\}$ which is also an independent set.
Since the commutation relations do not satisfy the standard particle creation (annihilation) algebra, we do not expect the quasi-particles for the problem to be electron like. We therefore define new fermionic quasi-particle operators that mix the electronic operators as,

\begin{eqnarray}\label{btransformation}
c_{{\bf k}\sigma l} = \sum'_n \left( u_{{\bf k}\sigma l}^{n}\gamma_n -\sigma v_{{\bf k}\sigma l}^{n*} \gamma_{n}^{\dagger} \right),
\end{eqnarray}

\noindent where $\sigma = \pm 1$ for $\uparrow\downarrow$ spin configuration. The prime on the summation is a restriction to include only those states in the summation that will lead to a positive energy excitation spectrum of the diagonal Hamiltonian. Using this transformation, the Hamiltonian is diagonalized in the following form, 

\begin{eqnarray}
H^{MF} &=& \sum_n E_n\gamma^{\dagger}_{n}\gamma_{n} + E_{\text{const}},
\end{eqnarray}

\noindent with,
\begin{eqnarray}\label{bcommutation}
\left[ \gamma_n,H^{MF} \right] &=& E_n\gamma_n, \nonumber \\
\left[ \gamma^{\dagger}_n,H^{MF} \right] &=& -E_n\gamma^{\dagger}_n.
\end{eqnarray}

\noindent Here, $E_n$ is the excitation spectrum for the Bogoliubov quasiparticles. The physical constraint of non-negativity on the excitation energies is implemented by discarding the negative energy states from the definition of the Bogoliubov transformation and hence the prime on the sum in Eq. (\ref{btransformation}). We now plug in our transformation Eq. (\ref{btransformation}) into the commutation relations Eq. (\ref{ecommutation}) using Eq. (\ref{bcommutation}) and we arrive at the BdG equations for the bilayer model. For convenience, the BdG equations can be recast in the matrix form as,
\begin{eqnarray}
\begin{pmatrix}
M_{4\times 4}{(\bf k)} & 0 \\ 
0 & \tilde{M}_{4\times 4}{(\bf k)}
\end{pmatrix} \Phi({\bf k}) = E_n\Phi({\bf k}),
\end{eqnarray}

\noindent with a symmetry that eigenspectrum of $M({\bf k})$ can be mapped to $\tilde{M}({\bf k})$ and vice versa without any loss of information. Therefore solving one of the $4\times 4$ block diagonals would suffice. The details to this approach can be found in \cite{Jian-xin2016}. We now restrict ourselves to the first block,

\begin{eqnarray}
M_{4\times 4}{(\bf k)} \Psi({\bf k}) = E_n\Psi({\bf k}),
\end{eqnarray}

\noindent where $\Psi^{T}({\bf k}) = \begin{pmatrix} u_{{\bf k}\uparrow 1} & v_{{\bf -k}\downarrow 1} & u_{{\bf k}\uparrow 2} & v_{{\bf -k}\downarrow 2} \end{pmatrix}$ and,

\begin{eqnarray}
M_{4\times 4}({\bf k}) = \begin{pmatrix}
						\xi_{1}({\bf k}) & \Delta_{1}^{\uparrow\downarrow}({\bf k}) & -\tilde{t} & 0 \\ 
						\Delta_{1}^{\uparrow\downarrow^{*}}({\bf k}) & -\xi_{1}({\bf -k}) & 0 & \tilde{t} \\
						-\tilde{t} & 0 & \xi_{2}({\bf k}) & 0 \\
						0 & \tilde{t} & 0 & -\xi_{2}({\bf -k})\\
						\end{pmatrix}.
\end{eqnarray}

\begin{figure}[H]
\centering
\renewcommand{\thefigure}{S\arabic{figure}}
\includegraphics[scale=0.32]{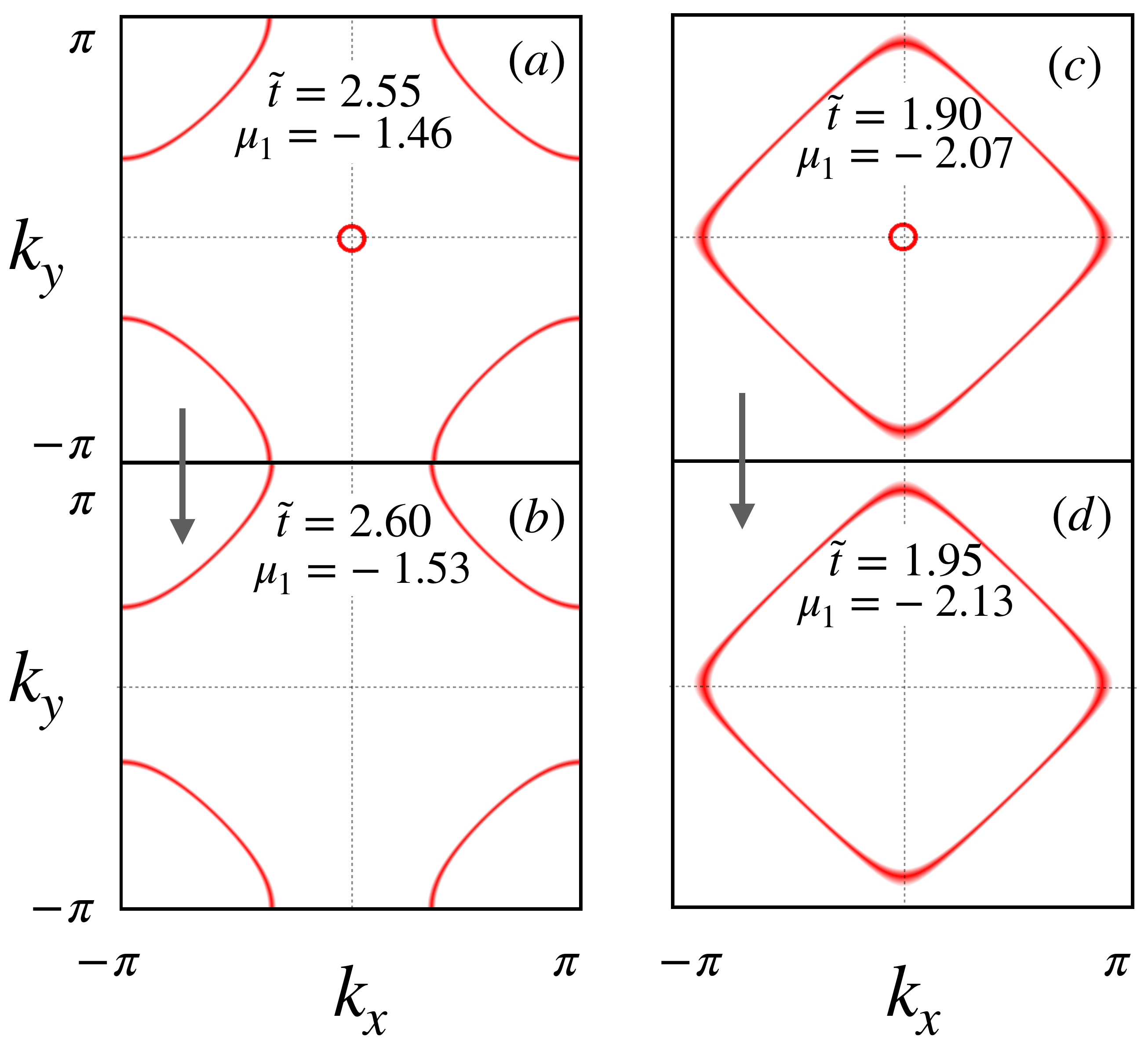}
\caption{Fermi surfaces of the non-interacting bilayer model for parameters that take us across two topological transitions reported in the main text. We keep $t_1=t_2=1$ and $\mu_1=\mu_2$ for all the plots. The specific choice of parameters is indicated in the plots, and corresponds to topological transition in Fig. 1($c$) and Fig. 1($d$) of the main text for panels ($a$)-($b$) and ($c$)-($d$), respectively. Note the disappearance of the small Fermi pockets near ${\bf k} = 0$ in going from panel ($a$) to panel ($b$), and also from ($c$) to ($d$).}
\label{fig:lifshitz}
\end{figure}

Notice the form of this Hamiltonian, the first $2\times 2$ block is a standard BdG Hamiltonian for the first layer and lower $2\times 2$ is just a metallic tight-binding Hamiltonian for the second layer. These two blocks are coupled via interlayer hopping $\tilde{t}$ appearing in the off-diagonal blocks. The transformation coefficients are the individual eigenvector components of this matrix and are used to calculate the quantum expectation values using Eq. (5) in the main text.

\noindent {\underline{\it Connection to Lifshitz transitions}:}
In this section we show that some of the transitions reported in the main text correlate perfectly with Lifshitz transitions in the corresponding non-interacting model. 
The Hamiltonian for the non-interacting model is defined as two tight-binding layers coupled via inter-layer hopping,
\begin{eqnarray}
H  &=&  -\sum_{\langle ij\rangle, \sigma, l} t_l \left( c_{i\sigma l}^{\dagger}c_{j\sigma l}+H.c.\right) - \sum_{i,\sigma,l} \mu_l c^{\dagger}_{i\sigma}c_{i\sigma} \nonumber \\
&& - \tilde{t} \sum_{i,\sigma} \left( c_{i\sigma 1}^{\dagger}c_{i\sigma 2}+H.c. \right).
\end{eqnarray}

\begin{figure}[]
\centering
\renewcommand{\thefigure}{S\arabic{figure}}
\includegraphics[scale=0.30]{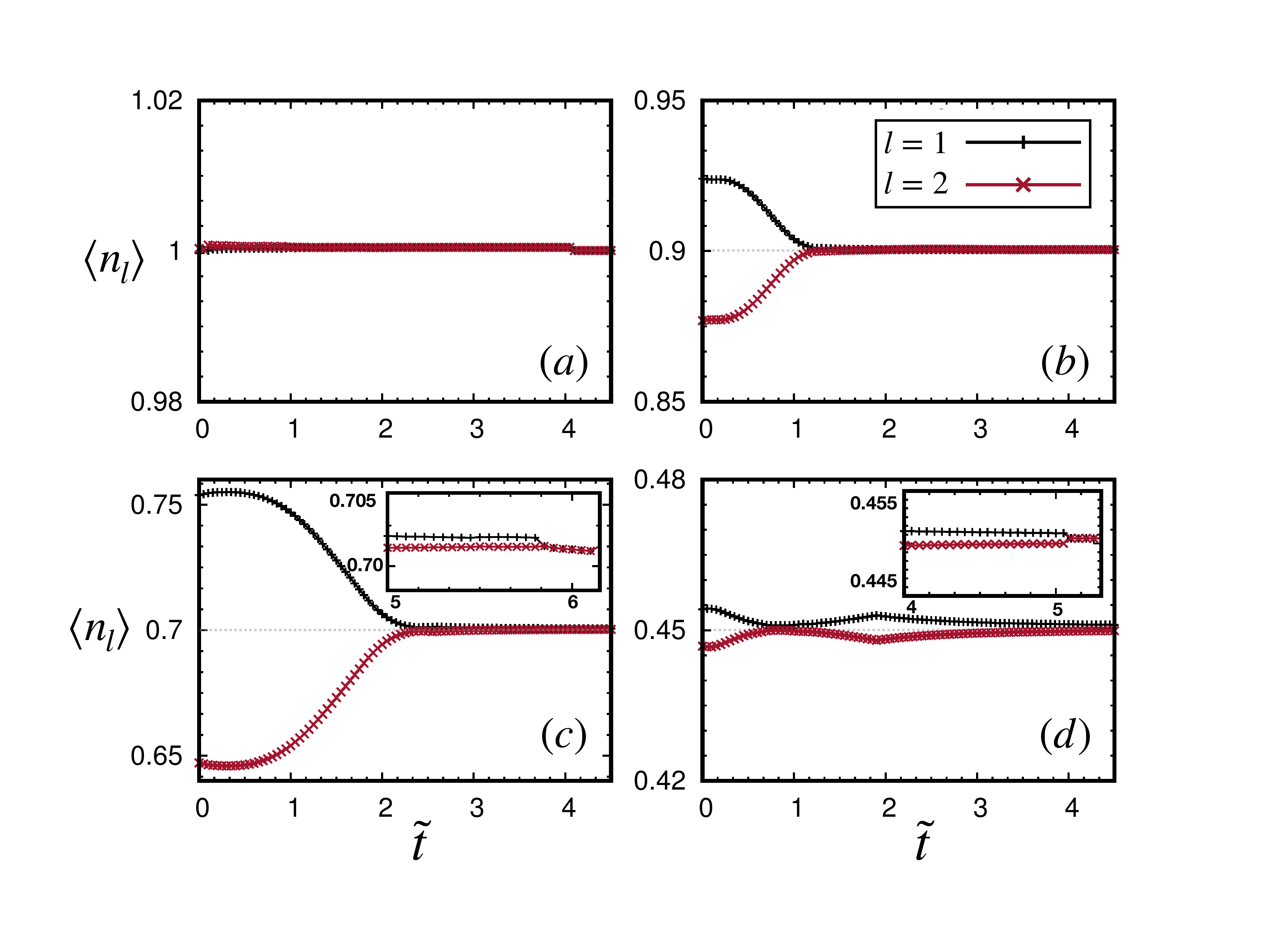}
\caption{($a$)-($d$) Layer-wise average electron density corresponding to Fig. 1($a$)-($d$), respectively, of the main text. Note that the density in two layers becomes equal for large ${\tilde t}$, as expected in the strong hybridization limit.}
\label{fig:density}
\end{figure}

\begin{figure}[]
\centering
\renewcommand{\thefigure}{S\arabic{figure}}
\includegraphics[scale=0.30]{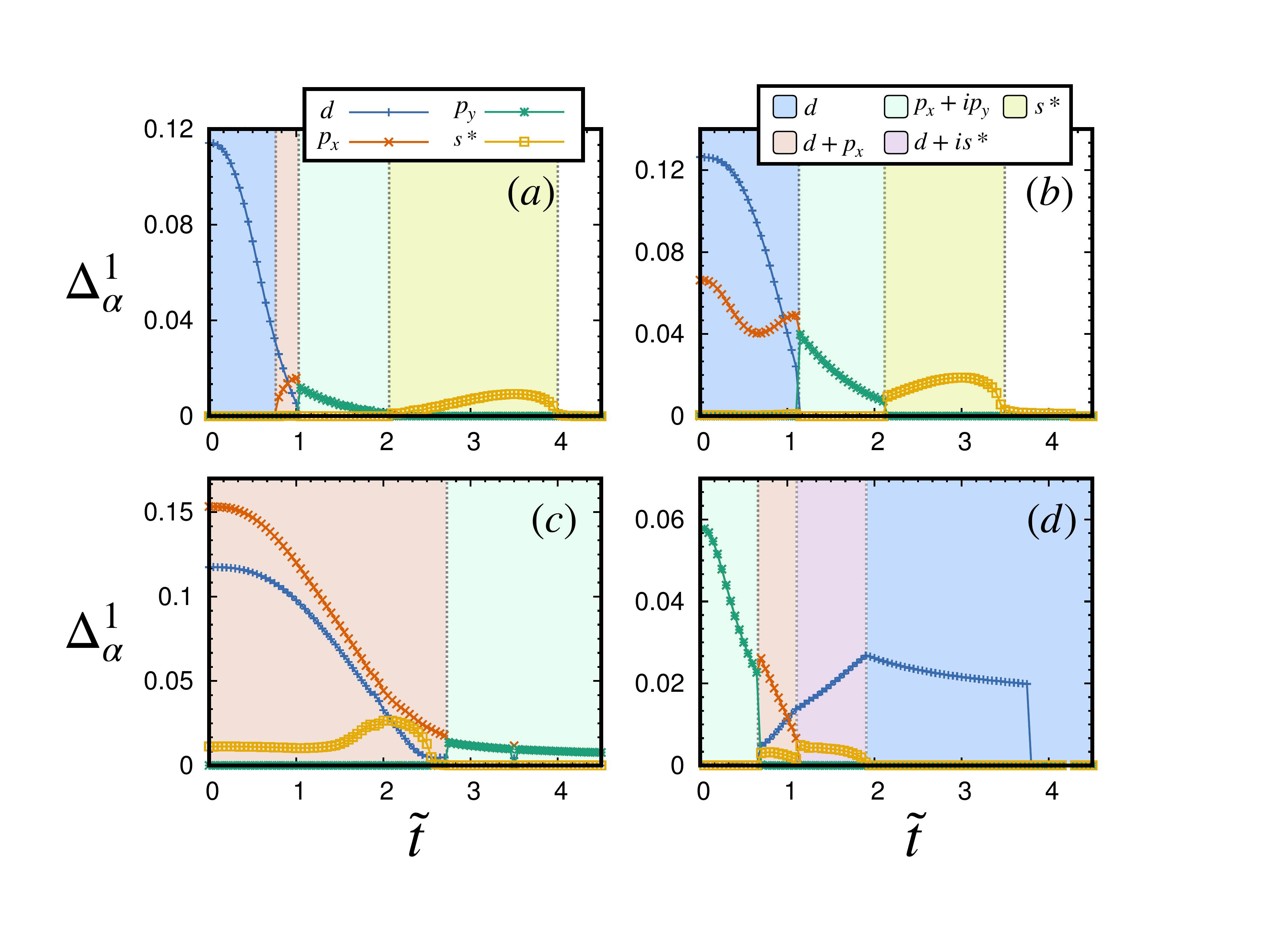}
\caption{Superconducting order parameter variation with $\tilde{t}$ when equal average density is enforced in each layer that requires $\mu_1\ne \mu_2$, in general. ($a$) $\langle n_1 \rangle = \langle n_2 \rangle = 1.0$, (b) $\langle n_1 \rangle = \langle n_2 \rangle = 0.9$, (c) $\langle n_1 \rangle = \langle n_2 \rangle = 0.7$, and (d) $\langle n_1 \rangle = \langle n_2 \rangle = 0.45$. All the remaining parameters are same as in Fig. 1 of the main text. Note that the various transitions shown in Fig. 1 of main text are also present here.}
\label{fig:tunnelingInduced}
\end{figure}

\noindent Notation is identical to that used in the main text. For model parameters corresponding to that used in Fig. 1 ($c$)-($d$) in the main text, we plot the Fermi surfaces of the non-interacting Hamiltonian in Fig. \ref{fig:lifshitz}. The change in parameter choice for panels ($a$) to ($b$) is such that the corresponding interacting system undergoes a transition from the $d+p_x+s*$ form of the order parameter to the topologically nontrivial $p_x+ip_y$ form captured in Fig. 1 ($c$) in the main text. The non-interacting Fermi surfaces show a qualitative change. The electron-like Fermi pocket near ${\bf k} = 0$ disappears. Therefore, one of the non-interacting bands stops contributing to pairing. A similar effect is demonstrated in Fig. \ref{fig:lifshitz} ($c$)-($d$), where the Lifshitz transition corresponds to the topological transition in Fig. 1($d$) in the main text. 
We emphasize that this correspondence between transitions in the form of the Superconducting order parameters and the Lifshitz transitions in the underlying non-interacting bands is not found for all the transitions \cite{Volovik2017}. However, we find another interesting connection between topological transitions and Lifshitz transitions on a $d-1$ dimensional metallic system living on the edges in cylinder geometry. This is briefly discussed in the main text.

\noindent {\underline{\it Effect of constant density in each layer}:}
In this section we show that the existence of topological transitions do not depend on the constraint ($\mu_1 = \mu_2$) on the layer dependent chemical potentials used in Fig. 1 of the main text. Firstly, we plot the change in layer-resolved electron density as a function of ${\tilde t}$ in Fig. \ref{fig:density}. This shows that for the choice of model parameters, the densities in the two layers are not very different from each other. In fact, for large values of ${\tilde t}$ one expects the hybridization to dominate and the densities in two layers should match. This is indeed obtained within our calculations, as shown in Fig. \ref{fig:density} ($a$)-($d$). Insets in Fig. \ref{fig:density} ($c$)-($d$)
display how the densities in two layers become identical for large ${\tilde t}$. These critical values of ${\tilde t}$ also mark the loss of superconductivity in the bilayer, which was not shown in the main text (Fig. 1($c$)-($d$) in main text).
Next, we work with an entirely different protocol to confirm the robustness of the transitions. We enforce equal density in each layer, which in general requires working with different values of the chemical potentials $\mu_1$ and $\mu_2$. The resulting variation of the different order parameter amplitudes is shown in Fig. \ref{fig:tunnelingInduced}. Comparing Fig. 1 of the main text with Fig. \ref{fig:tunnelingInduced}, it is clear that the qualitative features match very well and therefore, the results are insensitive to specific choice of the protocol of keeping densities or chemical potentials in each layer constant. One minor difference is the reduction in the critical value of ${\tilde t}$ required to kill the superconducting state (compare Fig. \ref{fig:tunnelingInduced} ($d$) and Fig. 1($d$) in main text). 
We would like to emphasis that this feature rules out the possibility that superconducting transitions reported in this work can simply be a consequence of effective density change in layer 1. The transitions as a function of $\mu_1 - \mu_2$ are already discussed in the main text.

\end{document}